\documentclass[twocolumn,prd,aps,showpacs,epsf,floats,amsfonts,amssymb,amsmath]{revtex4}

\newcommand{\be}{\begin{equation}}\newcommand{\ee}{\end{equation}}
\newcommand{\bea}{\begin{eqnarray}}\newcommand{\eea}{\end{eqnarray}}
\newcommand{\brr}{\begin{array}}\newcommand{\err}{\end{array}}
\newcommand{\bit}{\begin{itemize}}\newcommand{\eit}{\end{itemize}}
\newcommand{\ben}{\begin{enumerate}}\newcommand{\een}{\end{enumerate}}
\newcommand{\bib}{\bibitem}

\def\lan{\langle}
\def\lf{\left}

\def\non{\nonumber}
\def\ran{\rangle}

\def\ri{\right}

\def\al{\alpha}\def\bt{\beta}

\def\te{\theta}

\def\si{\sigma}

\def\1{{_{1}}}\def\2{{_{2}}}

\begin{document}

\title{ Neutrino mixing as a source of dark energy}

\author{A.Capolupo${}^{\flat}$, S.Capozziello${}^{\sharp}$, G.Vitiello${}^{\flat}$}

\vspace{2mm}

\address{${}^{\flat}$Dipartimento di  Fisica "E.R. Caianiello"
and INFN, Universit\`a di Salerno, I-84100 Salerno, Italy
\\ ${}^{\sharp}$ Dipartimento di Scienze Fisiche, Universit\`a di Napoli "Federico II" and INFN Sez. di Napoli,
Compl. Univ. Monte S. Angelo, Ed.N, Via Cinthia, I-80126 Napoli,
Italy}

\begin{abstract}
We show that the vacuum condensate due to neutrino mixing in
quantum field theory (QFT) contributes to the dark energy budget
of the universe which gives rise to the accelerated behavior of
cosmic flow. The mechanism of neutrino mixing is therefore 
a possible candidate to
contribute to the cosmological dark energy.
\end{abstract}


\maketitle

Data coming from cosmic microwave background radiation (CMBR)
\cite{CMBR,Spergel}, large scale structure \cite{LSS,Szalay} and
type Ia supernovae \cite{SNeIa}, used as standard candles,
independently support the picture that the today observed universe
can be consistently described as an accelerating Hubble fluid where
the contribution of dark energy component to the total matter-energy
density is $\Omega_{\Lambda}\simeq 0.7$. The big challenge is then
the one of explaining such a bulk of dark energy component.

On the other hand, in recent years great attention has been
devoted to the neutrino mixing phenomenon. The theoretical
investigation of the neutrino mixing, firstly proposed by
Pontecorvo \cite{Pontecorvo:1957cp}, has been pursed in depth
\cite{Bilenky:1978nj,Bilenky,Mohapatra:1991ng,Wolfenstein:1977ue,
Mikheyev,Giunti:1991ca}, and more recently the issue of the
construction of the flavored space of states has been settled in
the framework of the quantum field theory (QFT) formalism
\cite{BV95,BHV98,yBCV02,Capolupo:2004pt,Fujii:1999xa,JM011,Blasone:2005ae}
with the discovery of the unitary inequivalence between the
flavored vacuum and the massive neutrino vacuum \cite{BV95,BHV98},
the associated finding of the neutrino-antineutrino pair
condensate contributing to the vacuum energy \cite{Blasone:2004yh}
and the new oscillation formulas
\cite{BHV98,Fujii:1999xa,JM011,yBCV02}. The recent experimental
achievements proving neutrino oscillations \cite{SNO,K2K} and the
progresses in the QFT theoretical understanding
\cite{Blasone:2005ae} of the neutrino mixing thus provide a
challenging and promising path beyond the Standard Model of
electro-weak interaction for elementary particles.

In this paper we show that these two most interesting issues are
intimately bound together in such a way that one of them, namely the
neutrino mixing phenomenon, appears to provide a contribution, till
now unsuspected, to the vacuum dark energy component.  The structure of 
the flavor vacuum and its
unitary inequivalence to the vacuum for the massive neutrinos play a
significant r\^ole in obtaining the result we report below.

We want to stress that in this paper we do not solve the problem of the
arbitrariness of the  momentum
cut-off.  The  cut-off 
problem should therefore not be confused with the
main message we want to present in this paper, namely that the very
same phenomenon of the neutrino mixing appears to provide a
contribution to the cosmological dark energy component. As shown below,
such a feature, unsuspected till now, clearly manifests itself
provided the mathematically correct QFT treatment of neutrino mixing
\cite{BV95,BHV98,yBCV02,Capolupo:2004pt,Fujii:1999xa,JM011,Blasone:2005ae,Blasone:2004yh}
is considered.

In the simplest explanation, the so called $\Lambda CDM$ model,
the cosmological constant contributes for almost $70\%$ to the
total matter-energy density budget. The standard theory of
cosmological constant is based on the fact that the vacuum zero
point energy cannot violate the Lorentz invariance of the vacuum
and therefore the corresponding energy-momentum tensor density has
the form ${\cal T}_{\mu\nu}^{vac} = \lan 0 |{\cal T}_{\mu\nu}|
0\ran = \rho_{\Lambda} g_{\mu\nu}$, where $\rho_{\Lambda} $ is a
constant, i.e. a Lorentz scalar quantity. In the traditional
picture the vacuum itself can be thought as a perfect fluid,
source of the Einstein field equations and one derives
\cite{Sahni:2004ai} the equation of state $p_{\Lambda} = w
\rho_{\Lambda}$, with $p_{\Lambda}$ denoting the vacuum pressure
and the adiabatic index $w$ equals $-1$.
 As well known \cite{Itz}, one of the central
pillars of Lorentz invariant local QFT is the very same definition
of the vacuum state according to which it is the zero eigenvalue
eigenstate of the normal ordered energy, momentum and angular
momentum operators. Therefore, excluding by normal ordering
zero-point contributions, any non-vanishing vacuum expectation
value of one of these operators signals the breakdown of Lorentz
invariance, since the vacuum would be dependent on space and/or
time. The Lorentz invariance vacuum therefore implies ${\cal
T}_{\mu\nu}^{vac} = \lan 0 |:{\cal T}_{\mu\nu}:| 0\ran= 0$, (as
usual normal ordering is denoted by the colon $:...:$).

Usually, in the jargon one roughly expresses the Lorentz invariant
characterization of the QFT vacuum, by saying that preserving the
Lorentz invariance requires to exclude that kinematical terms in the
energy-momentum tensor may contribute to the vacuum expectation
values.

Suppose that, as said above, the  contribution of the (zero point)
vacuum energy density is taken to be equivalent to that of the
cosmological constant $\Lambda$, which is expressed by
$\rho_{\Lambda}=\Lambda/(8\pi G)$. Then, however, it turns out
that the vacuum expectation value of the energy-momentum tensor is
divergent, both for bosonic and fermionic fields, and this
shortcoming can be addressed as the {\it cosmological constant
problem}. By choosing to regularize the energy-momentum tensor by
an ultraviolet cut-off at Planck scale, one gets a huge value for
the vacuum energy density $\rho_{vac} \simeq c^5/G^2 \hbar\sim
10^{76} GeV^4$ which is 123 orders of magnitude larger than the
currently observed $\rho_{\Lambda}\simeq 10^{-47}GeV^4$
\cite{Zeldovich}. Also using a quantum chromo-dynamics (QCD)
cut-off \cite{shuryak} the problem is not solved since
$\rho_{\Lambda}^{QCD}\sim 10^{-3} GeV^4$ is still enormous with
respect to the actual observed value.

Furthermore, there is another aspect which has to be taken into
account: observations point out that cosmic flow is "today"
accelerating while it was not so at intermediate redshift $z$ (e.g.
$1 < z < 10$). This situation gave rise to structure formation
during the matter dominated era \cite{odintsov1,odintsov2}. This is
an indication of the fact that any realistic cosmological model
should roughly undergo four phases: an early accelerated phase
(inflation),  intermediate decelerated phases (radiation and matter
dominated) and a final, today observed, accelerated phase.
Obviously, the dynamical evolution (time dependence) of the
cosmological constant through the different phases, breaks the
Lorentz invariance of the vacuum and one has to face the problem of
the dynamics for the vacuum energy in order to match the
observations. In this case, therefore, we are not properly dealing
with cosmological constant. Rather, we have to take into account
some form of {\it dark energy} which {\it evolves} from early epochs
inducing the today observed acceleration. Such a dynamical evolution
of the dark energy, namely time dependence of the energy vacuum
expectation value, violates the Lorentz invariance.

In the literature there are many proposals to achieve cosmological
models justifying such a dark energy component, ranging from
quintessence \cite{Sahni:1999gb}, to braneworld
\cite{Deffayet:2001pu}, to extended theories of gravity
\cite{Capozziello:2005ku}. These approaches essentially consist in
adding new ingredients to the dynamics (e.g. scalar fields), or in
modifying cosmological equations (e.g. introducing higher order
curvature terms in the effective gravitational action).

In this letter we show that, due to the
condensate of neutrino-antineutrino pairs, the vacuum expectation
value of the energy-momentum tensor naturally provides a
contribution to the dark energy $\rho_{vac}^{mix}$, which in the
early universe satisfies the strong energy condition (SEC)
$\rho_{vac}^{mix} + 3 p_{vac}^{mix} \geq 0$, and at present epoch
behaves approximatively as a cosmological constant. Here
$p_{vac}^{mix}$ is the vacuum pressure induced by the neutrino
mixing. Under such a new perspective, the energy content of the
vacuum condensate could be substantially interpreted as
dynamically evolving dark energy.

The main features of the QFT formalism for the neutrino mixing are
summarized as follows. For the sake of simplicity, we restrict
ourselves to the two flavor case \cite{BV95}. Extension to three
flavors can be found in Ref. \cite{yBCV02}.
The relation between the Dirac flavored neutrino fields
$\nu_{e}(x)$, $\nu_{\mu}(x)$  and the Dirac massive neutrino
fields $\nu_{1}(x)$, $\nu_{2}(x)$ is given by
\bea \label{fermix1}
 \left(%
\begin{array}{c}
  \nu_{e}(x) \\
  \nu_{\mu}(x) \\
\end{array}%
\right)=\begin{pmatrix}
  \cos\theta & \sin\theta\\
  -\sin\theta & \cos\theta
\end{pmatrix}\left(%
\begin{array}{c}
  \nu_{1}(x) \\
  \nu_{2}(x) \\
\end{array}%
\right) \eea
being $\te$ the mixing angle.  The mixing transformation
(\ref{fermix1}) can be written as $\nu_{\si}(x)\equiv G^{-1}_{\bf
\te}(t) \; \nu_{i}(x)\; G_{\bf \te}(t),$
  where $(\si,i)=(e,1), (\mu,2)$,
and $G_{\bf \te}(t)$ is the transformation generator. The flavor
annihilation operators are defined as $\al^{r}_{{\bf k},\si}(t)
\equiv G^{-1}_{\bf \te}(t)\;\al^{r}_{{\bf k},i} \;G_{\bf \te}(t)$
 and $ \bt^{r}_{{-\bf k},\si}(t) \equiv
 G^{-1}_{\bf \te}(t)\; \bt^{r}_{{-\bf k},i}\;
G_{\bf \te}(t).$ They annihilate the flavor vacuum
$|0(t)\ran_{e,\mu}\,\equiv\,G_{\te}^{-1}(t)\;|0\ran_{1,2} $, where
$|0\ran_{1,2}$ is the vacuum annihilated by $\al^{r}_{{\bf k},i} $
and $ \bt^{r}_{{-\bf k},i}$.

The crucial point  of our discussion is that $|0(t)\ran_{e,\mu}$,
which is the physical vacuum where neutrino oscillations are
experimentally observed, is \cite{BV95}  a (coherent) condensate
of $\al_{{\bf k},i}$ ($\beta_{{\bf k},i}$) neutrinos
(antineutrinos):
\bea \non {}_{e,\mu}\langle 0| \al_{{\bf k},i}^{r \dag} \al^r_{{\bf
k},i} |0\rangle_{e,\mu}= {}_{e,\mu}\langle 0| \beta_{{\bf k},i}^{r
\dag} \beta^r_{{\bf k},i} |0\rangle_{e,\mu}= \sin^{2}\te ~ |V_{{\bf
k}}|^{2},\\ \label{con}\eea
where $i=1,2$, the reference frame ${\bf k}=(0,0,|{\bf k}|)$ has
been adopted for convenience, $V_{\bf k}$ is the Bogoliubov
coefficient entering the mixing transformation (see for example
Refs. \cite{BV95,yBCV02}) and  $|0\rangle_{e,\mu}$ denotes
$|0(t)\ran_{e,\mu}$ at a conventionally chosen time $t=0$. As a
consequence of its condensate structure the physical vacuum
$|0(t)\ran_{e,\mu}$ turns out to be unitary inequivalent to
$|0\ran_{1,2}$ \cite{BV95}. For brevity, we omit here to reproduce
the explicit expression of $V_{\bf k}$ which can be found, e.g.,
in Refs. \cite{BV95,yBCV02,Blasone:2005ae}. We only recall that
$V_{\bf k}$ is zero for $m_{1}= m_{2}$, it has a maximum at $|{\bf
k}|=\sqrt{m_\1 m_\2}$ and, for $ |{\bf k}| \gg\sqrt{m_\1 m_\2}$,
it goes like $|V_{{\bf k}}|^2\simeq (m_\2 -m_\1)^2/(4 |{\bf
k}|^2)$. For massless neutrinos, as well known, one does not have
mixing ($m_{1}= m_{2} = 0$). The oscillation formulas for the flavor charges
$Q_{e,\mu}(t)$ are obtained by computing their expectation values
in the physical vacuum $|0\rangle_{e,\mu}$
\cite{yBCV02,Blasone:2005ae}.

Let us now calculate the contribution $\rho_{vac}^{mix}$ of the
neutrino mixing to the vacuum energy density. We consider the
Minkowski metric (therefore we use the notation $\eta^{\mu\nu}$
instead of $g^{\mu\nu}$). The particle mixing and oscillations in
curved background will be analyzed in a separate paper. Eq.
(\ref{con}) suggests  that the energy content of the physical
vacuum  gets contributions from the $\al_{{\bf k},i}$ and
$\beta_{{\bf -k},i}$ neutrino condensate. Therefore, as customary
in such circumstances, we compute the total energy 
$T_{00}=\int d^{3}x {\cal T}_{00}(x)$ for
the fields $\nu_{1}$ and $\nu_{2}$, \bea\label{2b}
 :T_{(i)}^{00}: = \sum_{r}\int
d^{3}{\bf k}\, \omega_{k,i}\lf(\al_{{\bf k},i}^{r\dag} \al_{{\bf
k},i}^{r}+ \beta_{{\bf -k},i}^{r\dag}\beta_{{\bf -k},i}^{r}\ri),
\eea with $i=1,2$ and where $:...:$ denotes  the normal ordering
of the $\al_{{\bf k},i}$ and $\beta_{{\bf -k},i}$ operators. Note
that $T_{(i)}^{00}$ is time independent.

We remark that we have ${}_{e,\mu}\lan 0 |:T_{(i)}^{00}:|
0\ran_{e,\mu}={}_{e,\mu}\lan 0(t) |:T_{(i)}^{00}:|
0(t)\ran_{e,\mu} $, for any $t$, within the QFT formalism for
neutrino mixing. The contribution $\rho_{vac}^{mix}$ of the
neutrino mixing to the vacuum energy density is thus obtained:
 \bea
\frac{1}{V}\; {}_{e,\mu}\lan 0 |\sum_{i} :T_{(i)}^{00}:|
0\ran_{e,\mu} = \rho_{vac}^{mix}\; \eta^{00} ~.
 \eea

By using Eq.(\ref{con}), we then have
 \bea\label{cc}
\rho_{vac}^{mix} = \frac{2}{\pi} \sin^{2}\theta \int_{0}^{K} dk \,
k^{2}(\omega_{k,1}+\omega_{k,2}) |V_{\bf k}|^{2} , \eea
where the choice of the cut-off $K$ will be discussed below.

Similarly, the expectation value of $T_{(i)}^{jj}$ in the vacuum $|
0\ran_{e,\mu}$ gives the contribution
 $ p_{vac}^{mix}$ of the neutrino mixing
to the vacuum pressure:
 \bea\
\frac{1}{V}\; {}_{e,\mu}\lan 0 |\sum_{i} :T_{(i)}^{jj}:|
0\ran_{e,\mu} =  p_{vac}^{mix}\; \eta^{jj} ~,
 \eea
where no summation on the index $j$ is intended. Being, for each
diagonal component,
 \bea\label{5b} :T_{(i)}^{jj}:= \sum_{r}\int d^{3}{\bf k}\, \frac{k^j
k^j}{\;\omega_{k,i}}\lf(\al_{{\bf k},i}^{r\dag} \al_{{\bf
k},i}^{r}+ \beta_{{\bf -k},i}^{r\dag}\beta_{{\bf
-k},i}^{r}\ri),\eea (no summation on repeated indices), in the
case of the isotropy of the momenta: $k^{1}=k^{2}=k^{3}$, $3
(k^{j})^{2} = k^{2}$, we have $T^{11} = T^{22} = T^{33}$ and the
following equation holds
\bea\label{cc2}
 p_{vac}^{mix} &=& -\frac{2}{3\pi}
\sin^{2}\theta \int_{0}^{K} dk
k^{4}\lf[\frac{1}{\omega_{k,1}}+\frac{1}{\omega_{k,2}}\ri] |V_{\bf
k}|^{2}.
 \eea
Eqs.(\ref{cc}) and (\ref{cc2}) show that Lorentz invariance is
broken and
 $\rho_{vac}^{mix} \neq - p_{vac}^{mix}$ for any
value of the masses $m_1$ and $m_2$ and independently of the
choice of the cut-off. We observe that $w \simeq -1/3$ when the
cut-off is chosen to be $K \gg m_{1}, m_{2}$, cf. Eqs.(\ref{cc})
and (\ref{cc2}) and the discussion below for the choice of $K$.

It is worth stressing that the violation of the Lorentz invariance
originates from the neutrino-antineutrino condensate structure of
the vacuum. Indeed, as it appears from the computation reported
above, in the absence of such a condensate, i.e. with $|V_{\bf
k}|^{2} = 0$, the vacuum expectation value of {\it each} of the
$(0,0)$ and $(j,j)$ components of the energy-momentum tensor would
be zero. We also remark that the non-zero expectation value we
obtain is time-independent since, for simplicity, we are
considering the Minkowski metric. When the curved background
metric is considered, $|V_{\bf k}|^{2}$ gets a dependence on time,
as we will show in a forthcoming paper. In any case, the
contribution to the vacuum expectation value of $T^{\mu\nu}$ is
found to be non-vanishing, in the present computation (or in the
curved background case), not because the adopted metric is flat
(or not), but because of the non-trivial structure of physical
vacuum due to the mixing phenomenon (which manifests itself in the
non-vanishing of $|V_{\bf k}|^{2}$).

The above result holds in the early universe, when the universe
curvature radius is comparable with the oscillation length. At the
present epoch, in which the breaking of the Lorentz invariance is
negligible, the non-vanishing vacuum energy density
$\rho_{vac}^{mix}$ compatible with Lorentz invariance cannot come
from  condensate contributions carrying a non-vanishing
$\partial_{\mu}\sim k_{\mu}= (\omega_{k},k_{j})$, as it happens in
Eqs.(\ref{2b}) and (\ref{5b}) (see also Eqs.(\ref{cc}) and
(\ref{cc2})).
 This means that it can only be imputed to the lowest
energy contribution of the vacuum condensate, approximatively equal
to
\bea \rho_{\Lambda}^{mix}= \sum_{i}m_{i}\int
\frac{d^{3}x}{(2\pi)^3}\;{}_{e,\mu}\lan 0 |:\bar{\nu
}_{i}(x)\nu_{i}(x):| 0\ran_{e,\mu}.\eea
 Consistently with Lorentz
invariance, the state equation is now $\rho_{\Lambda}^{mix} \sim
-p_{\Lambda}^{mix}$, where explicitly
\bea\label{cost} \rho_{\Lambda}^{mix}=  \frac{2}{\pi}
\sin^{2}\theta \int_{0}^{K} dk \,
k^{2}\lf[\frac{m_{1}^{2}}{\omega_{k,1}}+\frac{m_{2}^{2}}
{\omega_{k,2}}\ri] |V_{\bf k}|^{2}. \eea

The result (\ref{cost}) shows that, at present epoch, the vacuum
condensate, coming from the neutrino mixing, can contribute to the
dark energy component of the universe, with a behavior similar to
that of the cosmological constant {\cite{Blasone:2004yh}.

We observe that, since, at present epoch, the characteristic
oscillation length of the neutrino is much smaller than the radius
of curvature of the universe, the mixing treatment in the flat
space-time, in such an epoch, is a good approximation of that in FRW
space-time. More interesting is also the fact that, at present
epoch, the space-time dependent condensate contributions, carrying a
non-vanishing $k_{\mu}$, are missing (they do not contribute to the
energy-momentum vacuum expectation value). The modes associated to
these missing contributions are not long-wave-length modes and
therefore they are negligible in the present flat universe, i.e with
respect to the scale implied by an infinite curvature radius.

Eqs. (\ref{cc}) and
(\ref{cost}) show that the contribution to the dark energy induced
from the neutrino mixing of course goes  to zero in the no-mixing
limit, i.e.  when the mixing angle $\theta = 0$ and/or $m_{1} =
m_{2}$, and clearly also for massless neutrinos.
However, those equations also show that the contribution
depends on the specific QFT nature of the mixing: indeed, it is
absent in the quantum mechanical (Pontecorvo) treatment, 
where $V_{\bf k}$ is anyhow zero. This confirms that the
contribution discussed above is a genuine QFT non-perturbative
feature  and it is thus of different origin with respect to the
ordinary vacuum energy contribution of massive spinor fields
arising from a radiative correction at some perturbative order
\cite{Coleman:1973}. This leads us to believe that a
neutrino--antineutrino asymmetry, if any, related with lepton
number violation \cite{Boyanovsky:2004xz}, would not affect much
our result. We will consider the problem of such an asymmetry in a
future work.

We call the reader attention on the fact that the discussion and the
related results presented till now, which constitute the core of our
message, are independent of the choice of the cut-off $K$. The
cut-off problem is a distinct problem to which we do not have a 
solution at the present.

As shown in Ref. \cite{Boyanovsky:2004xz}, in a dense background
of neutrinos,  as in the case of the early universe during the Big
Bang Nucleosyntesis, flavor particle-antiparticle pairs are
produced by mixing and oscillations with typical momentum $k \sim
\frac{m_{1} + m_{2}}{2}$, the average mass of the neutrinos. In this 
connection, we note that in the literature 
\cite{Blasone:2004yh,Barenboim:2004ev} it has been noticed that the
observed order of magnitude $\rho_{\Lambda}^{mix} \sim 
10^{-47}GeV^{4}$ can be reproduced by cutting the momentum range at the
QFT neutrino scales $\sqrt{m_{1} m_{2}}$ or $\frac{m_{1} + m_{2}}{2}$, with 
$\sin^{2}\theta\simeq 0.3$,  $m_{i}$ of order of $
10^{-3}eV$, so that $\delta m^{2}=m_{2}^{2}-m_{1}^{2}\simeq 8 \times
10^{-5}eV^{2}$. This reflects the presence of the $|V_{\bf
k}|^{2}$ factor, with its  momentum dependence, in the above
integrations, which points to the relevance of soft momentum 
(long-wave-length) modes.

In conclusion, we have shown that the vacuum condensate due to
neutrino mixing contributes to the dark energy budget of the
universe. Different behaviors of the vacuum expectation value of the
energy-momentum tensor have been discussed referring to different
boundary conditions in different universe epochs. Our result is independent of
the choice of the cut-off.

{\bf Acknowledgements}

We wish to thank Prof. T. W. Kibble for useful discussions and
encouragements. We acknowledge partial financial support from ESF
COSLAB Program, MIUR and INFN.

\bibliography{apssamp}

\end{document}